\documentclass[twocolumn,secnumarabic,amssymb, nobibnotes, aps, prx,superscriptaddress]{revtex4-2}

\usepackage{epstopdf}
\usepackage{xcolor}
\usepackage{ccaption}
\usepackage{graphicx}  
\usepackage{epstopdf}
\usepackage{dcolumn}   
\usepackage{bm}        
\usepackage{amssymb}   
\usepackage{amsmath}   
\usepackage{amsthm}
\usepackage{chemarrow}
\usepackage{color}
\usepackage{mathrsfs}
\usepackage{float}
\usepackage{bbold}
\usepackage{bbm}
\usepackage{textcomp}
\usepackage{url}
\usepackage{ccaption}
\usepackage{placeins}
\usepackage{lineno}
\makeatletter
\let\saved@includegraphics\includegraphics
\AtBeginDocument{\let\includegraphics\saved@includegraphics}
\renewenvironment*{figure}{\@float{figure}}{\end@float}
\makeatother

\begin{document}
\title{Implementation of a scalable universal two-qubit quantum processor \\  with electron and nuclear spins  in a trapped ion}

\author{Ji Bian}
\thanks{These authors contributed equally}
\affiliation{School of Physics and Astronomy, Sun Yat-Sen University, Zhuhai 519082, China}

\author{Teng Liu}
\thanks{These authors contributed equally}
\affiliation{School of Physics and Astronomy, Sun Yat-Sen University, Zhuhai 519082, China}

\author{Qifeng Lao}
\thanks{These authors contributed equally}
\affiliation{School of Physics and Astronomy, Sun Yat-Sen University, Zhuhai 519082, China}

\author{Min Ding}
\affiliation{School of Physics and Astronomy, Sun Yat-Sen University, Zhuhai 519082, China}

\author{Huiyi Zhang}
\affiliation{School of Physics and Astronomy, Sun Yat-Sen University, Zhuhai 519082, China}

\author{Xinxin Rao}
\affiliation{School of Physics and Astronomy, Sun Yat-Sen University, Zhuhai 519082, China}

\author{\\Pengfei Lu}
\affiliation{School of Physics and Astronomy, Sun Yat-Sen University, Zhuhai 519082, China}
\affiliation{Shenzhen Research Institute of Sun Yat-Sen University, Shenzhen 518057, China}
\affiliation{State Key Laboratory of Optoelectronic Materials and Technologies, Sun Yat-Sen University, Guangzhou 510275, China}

\author{Le Luo}
\email{luole5@mail.sysu.edu.cn}
\affiliation{School of Physics and Astronomy, Sun Yat-Sen University, Zhuhai 519082, China}
\affiliation{Shenzhen Research Institute of Sun Yat-Sen University, Shenzhen 518057, China}
\affiliation{State Key Laboratory of Optoelectronic Materials and Technologies, Sun Yat-Sen University, Guangzhou 510275, China}
\affiliation{Quantum Science Center of Guangdong-HongKong-Macao Greater Bay Area, Shenzhen 518048, China}


\date{\today}


\begin{abstract}
Increasing the quantum information processing power with limited number of hosts is vital for achieving quantum advantage.
Here we propose a novel scheme that achieves a scalable $n$-ion-$2n$-qubit quantum processor utilizing four internal levels of each ion, and experimentally implement a  $1$-ion-$2$-qubit universal processor using the valence electron spin and nuclear spin of a single $^{171}$Yb$^{+}$ ion. Fidelities of single-qubit and two-qubit gates are around $98\%$ obtained by quantum process tomography. Additionally, the Grover's algorithm is implemented with a successful rate exceeding $99\%$. We provide explicit scaling-up protocols based on standard laser-less and laser-based frameworks, and further demonstrate that the electron/nuclear-spin scheme allows less demanding two-qubit entangling gates between different ions.
The replacement of some inter-atomic gates by intra-atomic gates could increase the fidelity of some quantum circuits. Our work paves the way towards achieving $2^n$-times increase in the size of quantum computational Hilbert space with $n$ ions. 
%
\end{abstract}
\maketitle
\section{Introduction}
Quantum information processing (QIP) based on two-level qubits naturally inherits the binary information processing used in classical computers.
The physical hosts of quantum information, for example, atoms \cite{bruzewicz2019trapped,monroe2021programmable,saffman2010quantum,cai2023entangling} and superconducting circuits \cite{devoret2013superconducting,clarke2008superconducting,you2005superconducting}, however almost always consist of higher-dimensional Hilbert spaces. Meanwhile, increasing the QIP power with limited number of hosts and demonstrate quantum advantage remains a major goal in the  noisy intermediate-scale quantum (NISQ) era \cite{preskill2018quantum,arute2019quantum,zhong2020quantum}. These has stimulated studies on qudit-based QIP, i.e., using $d > 2$ levels of the underlying system as a QIP unit \cite{wang2020qudits}, with recent experimental progress in constructing single- and two-qudit processors \cite{ringbauer2022universal,hrmo2023native,liu2023performing}. 
While the d-level structure makes it more suitable for applications that are naturally formulated in high-dimensional Hilbert spaces \cite{macdonell2021analog,rico2018so}, it comes at a cost. First, 
it increases complexity in algorithms: While qudits offer the potential for increased computational power, developing algorithms that effectively utilize this additional dimensionality can be challenging. Secondly,
quantum algorithms and protocols designed for qubits may not directly translate to qudits. This can slow down progress in utilizing qudits for quantum information processing, as it requires additional research and development effort to adapt existing algorithms or develop new ones, as an example, standard qubit-based quantum error correction could not be directly implemented on qudit processors. Instead, people are looking forward to the scheme using $d > 2$ levels to realize a multi-qubit quantum processor in a single host \cite{campbell2022polyqubit,yang2022realizing}.
  
Here, we propose a novel scheme that utilizes the $4$-levels of a single trapped ion as \emph{two qubits}, and construct a scalable $2n$-qubit processor with $n$ ions. The four levels $|1'\sim 4'\rangle$ could be chosen arbitrarily (termed ``logical encoding''), e.g., $|1'\rangle=\left|\uparrow\uparrow\right\rangle$, $|2'\rangle=\left|\uparrow\downarrow\right\rangle$, $|3'\rangle=\left|\downarrow\uparrow\right\rangle$, $|4'\rangle=\left|\downarrow\downarrow\right\rangle$, where $\left|\uparrow,\downarrow\right\rangle$ represents encoded qubits (omitting the tensor product symbol). This is also proposed in recent works \cite{campbell2022polyqubit,simm2023two}. Or, as explicitly studied in this work, the two qubits are specifically represented by two physical $2$-level quantum systems (termed ``physical encoding''), of the electron spin and nuclear spin (subspace). This way of encoding is implicitly used in color center-based QIP \cite{prawer2008diamond,wu2019programmable,castelletto2020silicon}, but has not been applied to atom-based QIP experiments. 

We experimentally realize a 
$1$-ion-$2$-qubit processor using the valence electron spin, nuclear spin (spin-$1/2$) and their mutal hyperfine interaction  within a single $^{171}$Yb$^{+}$ ion under standard quantization magnetic fields \cite{debnath2016demonstration}, and implement high-fidelity universal two-qubit gates. We then perform Grover's algorithm as an application. We also give explicit scaling-up protocols based on both laser-less \cite{mintert2001ion,ospelkaus2008trapped,sutherland2019versatile,ospelkaus2011microwave,srinivas2021high} and laser-based frameworks \cite{wang2022fast,wong2017demonstration,schafer2018fast,de2016parallel}. We then discuss the readout process, which could be achieved using well-known techniques \cite{low2020practical,baldwin2021high}. We discover that the large discrepancy between electron (nuclear) gyromagnetic ratio $\gamma_e$ ($\gamma_n$) allows less demanding two-qubit entangling gates between different ions under large quantization magnetic fields \cite{wang2011quantum,feng2009nuclear}.

The main distinction between the $n$-ion-$2n$-qubit processor and a ($d=4$) qudit processor is as follows. Although both methods offers $2^{n}$-times increase in the size of QIP Hilbert space, in the $n$-ion-$2n$-qubit processor, individual qubit state preparation and readout in a single ion could be accomplished, without destroying the other qubit in the same ion. This allows standard quantum algorithms including quantum error correction to be implemented without modification, whereas this is not the case for qudits. Besides, the replacement of some inter-atomic gates (fidelity limited by motional state decoherence \cite{brownnutt2015ion}) by intra-atomic gates in our scheme could increase fidelities of certain quantum circuits. 

The logical encoding and physical encoding schemes each has their special features: the logical encoding could use four clock states, and thus offers relatively long coherence times. But it does not have a clear entanglement structure, for example, if the four levels are chosen from the spin angular momentum eigenstates of the same spin, then there is no entanglement at all. Nevertheless, as is also pointed out in \cite{campbell2022polyqubit}, most quantum computing algorithms do not require the nonlocality of quantum entanglement, and could therefore obtain the increased processing power within this scheme. The physical encoding scheme has well defined entanglement structure, and the identification of nuclear/electron spins enables certain tasks such as the quantum simulation of non-Abelian lattice gauge theories \cite{banerjee2013atomic}, where the internal symmetry of nuclear spins could be used to maintain gauge invariance. The naturally existing entanglement structure in this scheme, e.g., strong coupling between two kinds of qubits in individual ions plus adjustable coupling between different ions, also makes it advantageous in certain applications, such as the preparation of thermofield double states \cite{cottrell2019build,zhu2020generation}, and studying the dynamics of many-body scars \cite{serbyn2021quantum,zhang2023many}. 
The number of qubits and the dimension of QIP Hilbert space are not limited to $2$ and $4$ in a single ion: $2^p$ levels could be used to encode $p>2$ qubits. The higher angular momentum  ($>1/2$) structure (and its subspace) could be explored to construct hybrid qubit-qudit coupled systems, such as the $^{25}$Mg$^{+}$ ($^{43}$Ca$^{+}$) ion with nuclear spin $5/2$ ($7/2$), to simulate interacting spin ($>1/2$) systems.
Through viewing the information hosts from a different angle and exploring their underlying physical structures,  the $n$-ion-multi-qubit scheme would become a complement to the earlier developed qudit scheme on the way toward increasing the QIP power.

\section{Electron and nuclear spin as two qubits in a single trapped ion}\label{result}
\begin{figure}[htb]  
	\makeatletter
	\def\@captype{figure}
	\makeatother
	\centering
	\includegraphics[scale=0.98
	]{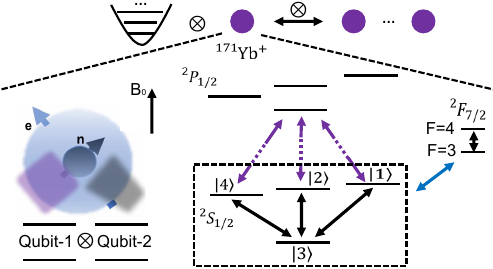}
	\caption{The $n$-ion-$2n$-qubit processor. The hyperfine interaction between nuclear spin (qubit-1) and electron spin (qubit-2) results in the $^{2}$S$_{1/2}$ 4-level subspace. The $^{2}$F$_{7/2}$ subspace could also be utilized as qubit levels and to perform readout.}
	\label{fig1}
\end{figure}

Here we take $^{171}$Yb$^{+}$ ion as an example, and demonstrate the physical encoding scheme (see fig.\ref{fig1}).
Consider the s-electron (spin-$1/2$) in the outermost shell and the $^{171}$Yb nucleus (spin-$1/2$). Denote the angular momentum operators $I_x,I_y,I_z$ as the standard Pauli matrices times $1/2$,
and eigenstates of $I_z$ as $\left|\uparrow \right\rangle$, $\left|\downarrow \right\rangle$, with corresponding eigenvalues $\hbar/2$, $-\hbar/2$ ( $\hbar=1$ in the following).
The magnetic dipole-dipole interaction between  these two spins results in the hyperfine interaction \cite{foot2004atomic}
$
H_A=A(I_{1,x}I_{2,x}+I_{1,y}I_{2,y}+I_{1,z}I_{2,z}),
$
where $A\approx 2\pi \times 12.6$ GHz is the hyperfine interaction constant, $1$ represents nucleus and $2$ electron. 

The ion is typically in an external magnetic field $B_0$ providing the quantization axis (denote it $z$ axis), so the free evolution Hamiltonian is
\begin{equation}\label{h00}
\begin{aligned}
H_0&=A(I_{1,x}I_{2,x}+I_{1,y}I_{2,y}+I_{1,z}I_{2,z})
-B_0(\gamma_1I_{1,z}+\gamma_2I_{2,z})\\
&=E_1|1\rangle\langle 1|+E_2|2\rangle\langle 2|+E_3|3\rangle\langle 3|+E_4
|4\rangle\langle 4|,
\end{aligned}
\end{equation}
\\
where  $\gamma_1/(2\pi)\approx 7.5 \times 10^{6}$ Hz$\cdot$T$^{-1}$$\cdot$s$^{-1}$, $\gamma_2 /(2\pi)\approx -2.8 \times 10^{10}$ Hz$\cdot$T$^{-1}$$\cdot$s$^{-1}$  are the gyromagnetic ratios. $|1\rangle$, $|2\rangle$, $|3\rangle$, $|4\rangle$ are eigenstates of $H_0$ with eigenvalues $E_{1\sim 4}$, they are related to $\left|\uparrow \uparrow\right\rangle$, $\left|\uparrow \downarrow\right\rangle$, $\left|\downarrow \uparrow\right\rangle$, $\left|\downarrow \downarrow\right\rangle$ by
\begin{equation}
\begin{aligned}
&|1\rangle=\left|\uparrow \uparrow\right\rangle,\quad |2\rangle=\cos\frac{\theta_0}{2}\left|\uparrow \downarrow\right\rangle-\sin\frac{\theta_0}{2}\left|\downarrow \uparrow\right\rangle, \\
&|3\rangle=-\sin\frac{\theta_0}{2}\left|\uparrow \downarrow\right\rangle-\cos\frac{\theta_0}{2}\left|\downarrow \uparrow\right\rangle,\quad |4\rangle=\left|\downarrow \downarrow\right\rangle,
\end{aligned}
\end{equation}
\\
where $\theta_0$ depends on $B_0$ ($\theta_0 \approx -\frac{\pi}{2}$ here).
Define the ``spin basis'' where $\left|\uparrow \uparrow \right\rangle_{\textrm{s}}=[1,0,0,0]^{\textrm{T}}$, $\left|\uparrow \downarrow \right\rangle_{\textrm{s}}=[0,1,0,0]^{\textrm{T}}$, $\left|\downarrow \uparrow \right\rangle_{\textrm{s}}=[0,0,1,0]^{\textrm{T}}$, $\left|\downarrow \downarrow \right\rangle_{\textrm{s}}=[0,0,0,1]^{\textrm{T}}$, the subscript indicates spin basis, and it is also the computational basis of the two-qubit processor.
Also define the ``number basis'' where $|1\rangle_{\textrm{n}}=[1,0,0,0]^{\textrm{T}}$, $|2\rangle_{\textrm{n}}=[0,1,0,0]^{\textrm{T}}$, $|3\rangle_{\textrm{n}}=[0,0,1,0]^{\textrm{T}}$, $|4\rangle_{\textrm{n}}=[0,0,0,1]^{\textrm{T}}$,
the subscript indicates number basis.
Define a mapping operator $R$ (depends on $B_0$) satisfying
\begin{equation}
\begin{aligned}
R|k\rangle_{\textrm{n}}=|k\rangle_{\textrm{s}} \quad (k=1,2,3,4),
\end{aligned}
\end{equation}
where $|k\rangle_{\textrm{s}}$ represents the matrix form of $|k\rangle$ in spin basis,
i.e., it maps the matrix form of a vector (or an operator) in number basis to that in spin basis, and vice versa.
The explicit form of $R$ reads
\begin{equation}\label{mapr}
R=
\begin{pmatrix} 1 & 0 & 0 & 0 \cr 0 & \cos(\theta_0/2) & -\sin(\theta_0/2) & 0 \cr 0 & -\sin(\theta_0/2) & -
\cos(\theta_0/2) & 0 \cr 0 & 0 & 0 & 1 \end{pmatrix},
\end{equation}
where
$\theta_0=2 \textrm{tan}^{-1}(\lambda)$ and
\begin{equation}
\lambda=\frac{-B_0\gamma_1+B_0\gamma_2-\sqrt{A^2+B^2_0\gamma^2_1+B^2_0\gamma^2_2-2B^2_{0}\gamma_1\gamma_2}}{A}.
\end{equation}
Eq. \eqref{h00} corresponds to the $^{2}S_{1/2}$ four-level subspace in $^{171}$Yb$^{+}$, as shown in fig.\ref{fig1}. It is clear that we could treat spin $1$, $2$ as two qubits with coupling, and $\left|\uparrow,\downarrow \right\rangle$ as the two levels of a qubit. 

As Zeeman sublevels are used, the coherence time is degraded compared to conventional clock-states based qubits. This could be circumvented by multilevel dynamical decoupling \cite{yuan2022preserving} or magnetic shielding \cite{ruster2016long}. The logical encoding scheme also works in these four levels, and could potentially achieve better coherence time if the $^{2}$S$_{1/2}$ $|F=0,1,m_F=0\rangle$ and $^{2}$F$_{7/2}$ $|F=3,4,m_F=0\rangle$ (lifetime $\approx 1.58$ yrs) clock states are utilized, at the expense of more complicated optical set-ups \cite{campbell2022polyqubit,yang2022realizing}.

\section{Universal quantum control in the $1$-ion-$2$-qubit system}\label{onetwo}
To demonstrate the universality of the $1$-ion-$2$-qubit processor, we choose to implement the following operations that form a universal quantum gate set \cite{nielsen2002quantum}: Hadamard, Phase, $\pi/8$, and CNOT gates.

There is a number of ways to implement these operations. For example, one could use analytical method to solve the Schr\"{o}dinger equation and obtain the control pulses that achieve the gate set. These pulses require simultaneous driving of e.g., $|3\rangle \leftrightarrow |1\rangle$ and $|2\rangle \leftrightarrow |4\rangle$, without driving  $|1\rangle \leftrightarrow |2\rangle$. This would require a large $B_0$ (to magnify the unequal splitting between $|2\rangle \leftrightarrow |1,4\rangle$), and therefore requires modification of current set-ups ($B_0 \approx 6$ Gs). Alternatively, one could even go to regimes where $|B_0\gamma_{1,2}| \gg A$. Single-qubit gates could now be realized by selective excitation based on different Larmor frequencies ($\gamma_{1,2}B_0$), and the secular-approximated coupling term $I_{1,z}I_{2,z}$ could be used to realize two-qubit gates \cite{vandersypen2005nmr}. This is similar to those applied in magnetic resonance experiments, but requires even larger magnetic fields and is therefore not applicable on current set-ups either \cite{wang2011quantum,feng2009nuclear}. There is yet another possibility, inspired by universal quantum control of spin systems in ultra-low magnetic field \cite{bian2017universal,jiang2018numerical,jiang2018experimental,albertini2002lie}, that using shaped dc-magnetic fields (without the $\approx 12.6$ GHz carrier frequencies) and standard quantization fields only, the universal gate set could still be realized. This is directly applicable on current trapped-ion setups and is worth further experimental studies.

Numerical optimal control is powerful in obtaining the desired control pulses \cite{glaser2015training}, it is also applicable in the current situation and is utilized in our experiment. Consider a control scenario applicable to common trapped-ion setups: three near-resonant control fields applied simultaneously to each ion. The control fields could be realized by microwaves, such as in our experiments among others \cite{lu2024experimental,bian2023quantum,bian2023implementation,lu2024realizing,zhao2022efficient,srinivas2021high,zhang2015time,hu2018experimental}. Or, they could be realized by laser fields \cite{debnath2016demonstration,zhang2017observation}. Transforming to a rotating frame with frequencies determined by the control fields and apply the rotating wave approximation (RWA) (see Appendix A), we get the total Hamiltonian 
\begin{equation}\label{hcrwa}
\begin{aligned}
H_c=&c_{31}(t)|3\rangle\langle 1|+c_{32}(t)|3\rangle\langle 2|+c_{34}(t)|3\rangle\langle 4|\\
&+\frac{\delta_{1}}{2}(t)|1\rangle \langle 1|+\frac{\delta_{2}(t)}{2}|2\rangle \langle 2|+\frac{\delta_{4}(t)}{2}|4\rangle \langle 4|+\textrm{H.c.},
\end{aligned}
\end{equation}
\\
where $c_{31,32,34}$ ($\delta_{1,2,4}$) are complex (real) numbers, and are determined by the strength and phase (detunings) of the near-resonant control fields.  

Denote $F$ the performance function 
that characterizes the fidelity of the implemented operation $U$ relative to the ideal $U_{\textrm{id}}$, the optimization task is therefore

\vspace{\baselineskip}
\hspace{6em} find \{$c_{31,32,34}(t)$, $\delta_{1,2,4}(t)$\},


\hspace{6em} max $F(U)$,


\hspace{6em} s.t. $\frac{\partial U(t)}{\partial t}=-iH_c(t)U(t)$.
\vspace{\baselineskip}

\begin{figure}[htb]  
	\makeatletter
	\def\@captype{figure}
	\makeatother
	\centering
	\includegraphics[scale=0.62
	]{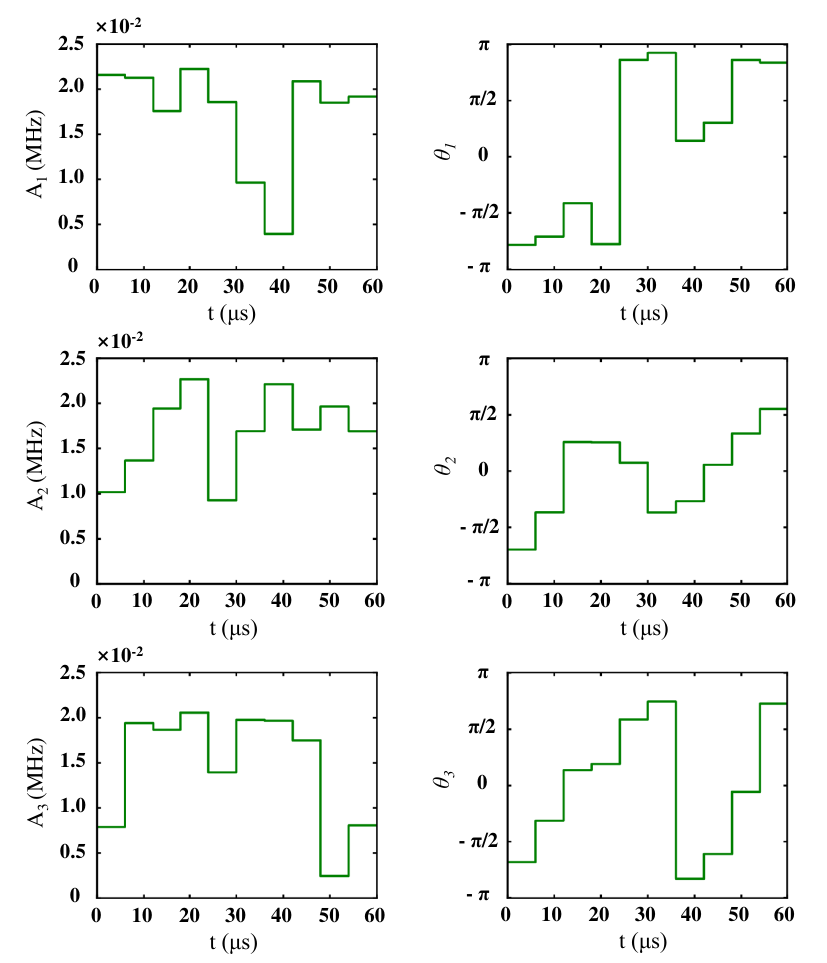}
	\caption{A theoretical pulse sequence achieving C$_{1}$NOT$_2$. $A_{1,2,3}$ and $\theta_{1,2,3}$ are corresponding amplitudes and phases of $c_{31,32,34}$}
	\label{grapetheory}
\end{figure}

\noindent While there are many quantum optimal control algorithms developed in these years \cite{machnes2018tunable,doria2011optimal,green2013arbitrary}, we use the Gradient Ascent Pulse  Engineering (GRAPE) algorithm \cite{khaneja2005optimal}, as it behaves well enough in accomplishing our tasks. We obtain control pulses achieving all the operations in the gate set with theoretical fidelities $>0.999$. As an example, a pulse sequence achieving C$_{1}$NOT$_2$ is illustrated in fig. \ref{grapetheory}. Here only $c_{31,32,34}$, with corresponding amplitudes $A_{1,2,3}$ and phases $\theta_{1,2,3}$ are modulated, $\delta_{1,2,4}(t)$ are kept zero. 

%

Quantum state tomography (QST) is required to obtain the full information of the state after an operation, and could be combined to implement the quantum process tomography (QPT) that enables  characterization of the performance of the implemented operation. In our experiment, the direct observable is $|3\rangle \langle 3|$. Populations on $|1\sim 4\rangle$ and coherence among them could be obtained by first transferring them to the $|3,1\rangle$, $|3,2\rangle$ or $|3,4\rangle$ subspace by $\pi$ and $\pi/2$ pulses, and then apply the standard qubit QST. The obtained $\tilde{\rho}$ is in the number basis, applying $R$ we obtain $\rho=R\tilde{\rho}R^{\dagger}$ in the spin basis, which is also the natural computational basis for the two-qubit processor.

\begin{figure}
	\centering
	\includegraphics[width=0.48\textwidth]{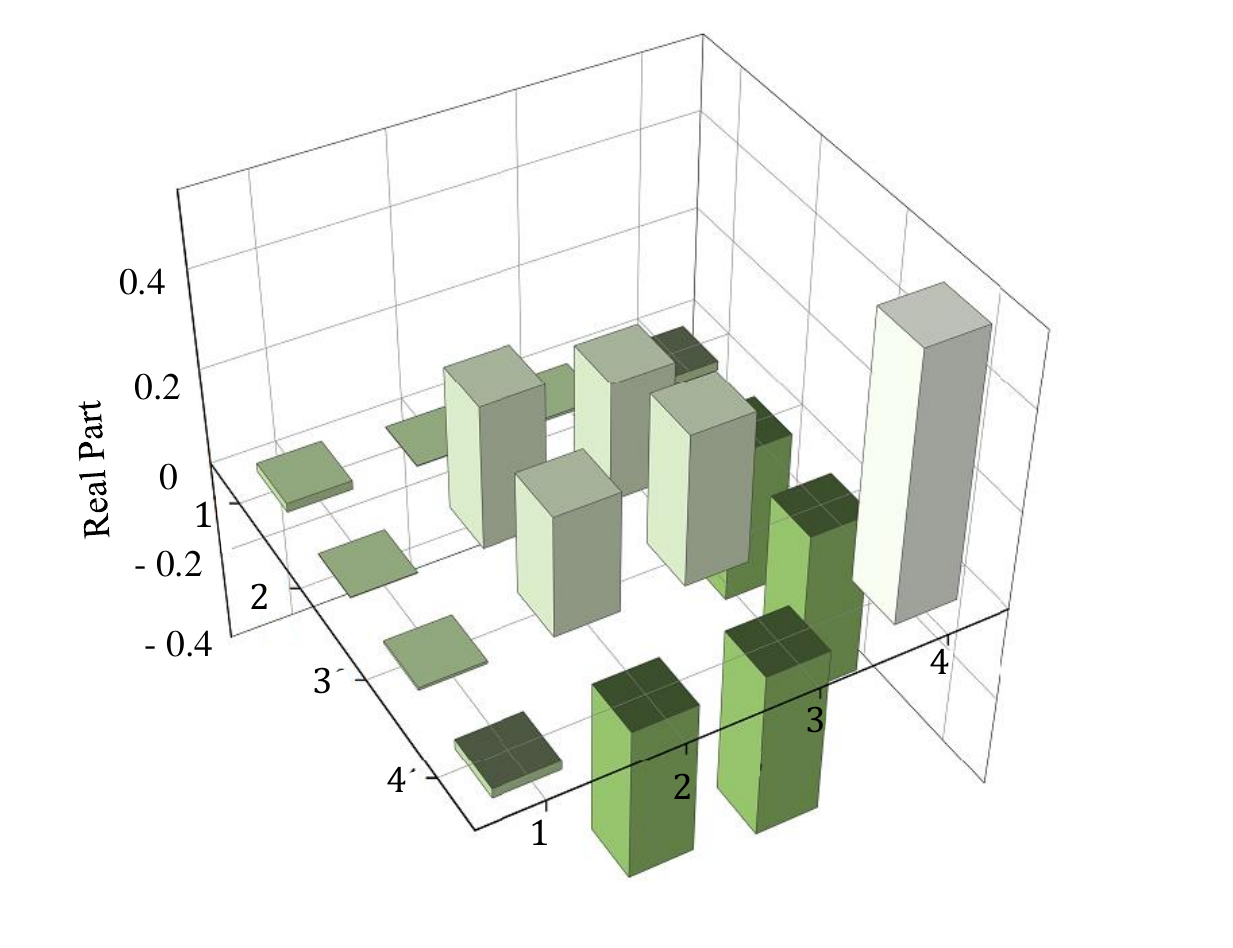}
	\caption{Real parts of the final state density matrix in the number basis after C$_1$NOT$_2$ Gate. Theoretical imaginary parts equal zero, absolute values of the measured imaginary parts are all smaller than $0.03$ and are not shown.}
	\label{C1NOT2_Re}
\end{figure}

The experiment is performed on the trapped $^{171}$Yb$^{+}$ ion quantum information processor (details listed in Appendix B). 
Microwave fields near resonant with the three transitions $|3\rangle \leftrightarrow |1,2,4\rangle$, with tunable frequencies, amplitudes and phases are used to control the spin states. The population $P_{3}$ on $|3\rangle$  could be readout by first measuring the total population $P_1+P_2+P_4$ on $|1\rangle$, $|2\rangle$, and $|4\rangle$ through applying the $369.5$ nm detection beam and measuring the fluorescence, then $P_{3}=1-(P_1+P_2+P_4)$. The initial state of each experiment is $|3\rangle$ prepared by optical pumping.
Starting from $|3\rangle$, the quantum states after application of operations in the gate set are obtained by QST. For example, the measured density matrix after C$_{1}$NOT$_{2}$ is shown in fig. \ref{C1NOT2_Re}. It is demonstrated in the number basis, and could be easily transferred to the spin basis by $R$. The state fidelities are listed in table \ref{tab:fidelity1}. We also implement the SWAP gate which is useful to transfer quantum information among different types of qubits.

\begin{table*}[htb]
	\caption{QST fidelity.}
	\begin{tabular}{ccccccccccc}
		\hline \hline 
	& Gate & C$_1$NOT$_2$ & C$_2$NOT$_1$ & Phase$_1$ & Phase$_2$ & $\pi/8_1$ & $\pi/8_2$ & Hadamard$_1$ & Hadamard$_2$ & SWAP \\
	\hline 
	 & Fidelity (\%) & 99.0 $\pm$0.19 & 99.0$\pm$0.45 & 99.0$\pm$0.51 & 99.0$\pm$0.14 & $98.9\pm0.33$ & $98.3\pm0.69$ & $98.4\pm0.43$ & $98.2\pm0.65$ & $99.7\pm0.14$  \\
		\hline \hline
	\end{tabular}\label{tab:fidelity1}
\end{table*}

\begin{figure}
	\centering
	\includegraphics[width=0.48\textwidth]{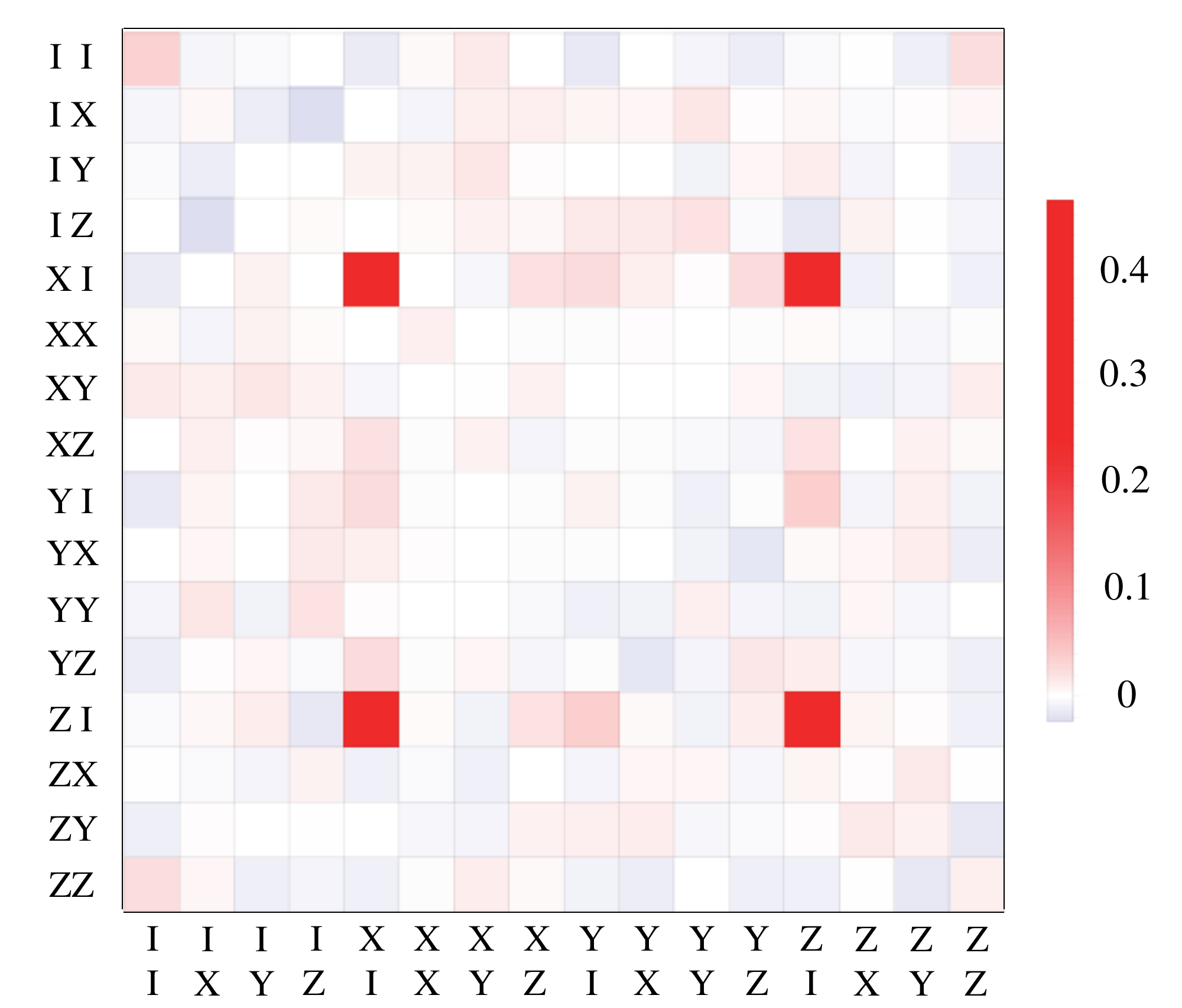}
	\caption{Real parts of the $\chi$ matrix of the Hadamard$_1$ Gate in spin basis. Theoretical imaginary parts equal zero, absolute values of the measured imaginary parts are all smaller than $0.02$ and are not shown.}
	\label{H1chi}
\end{figure}

\begin{figure}
	\centering
	\includegraphics[width=0.48\textwidth]{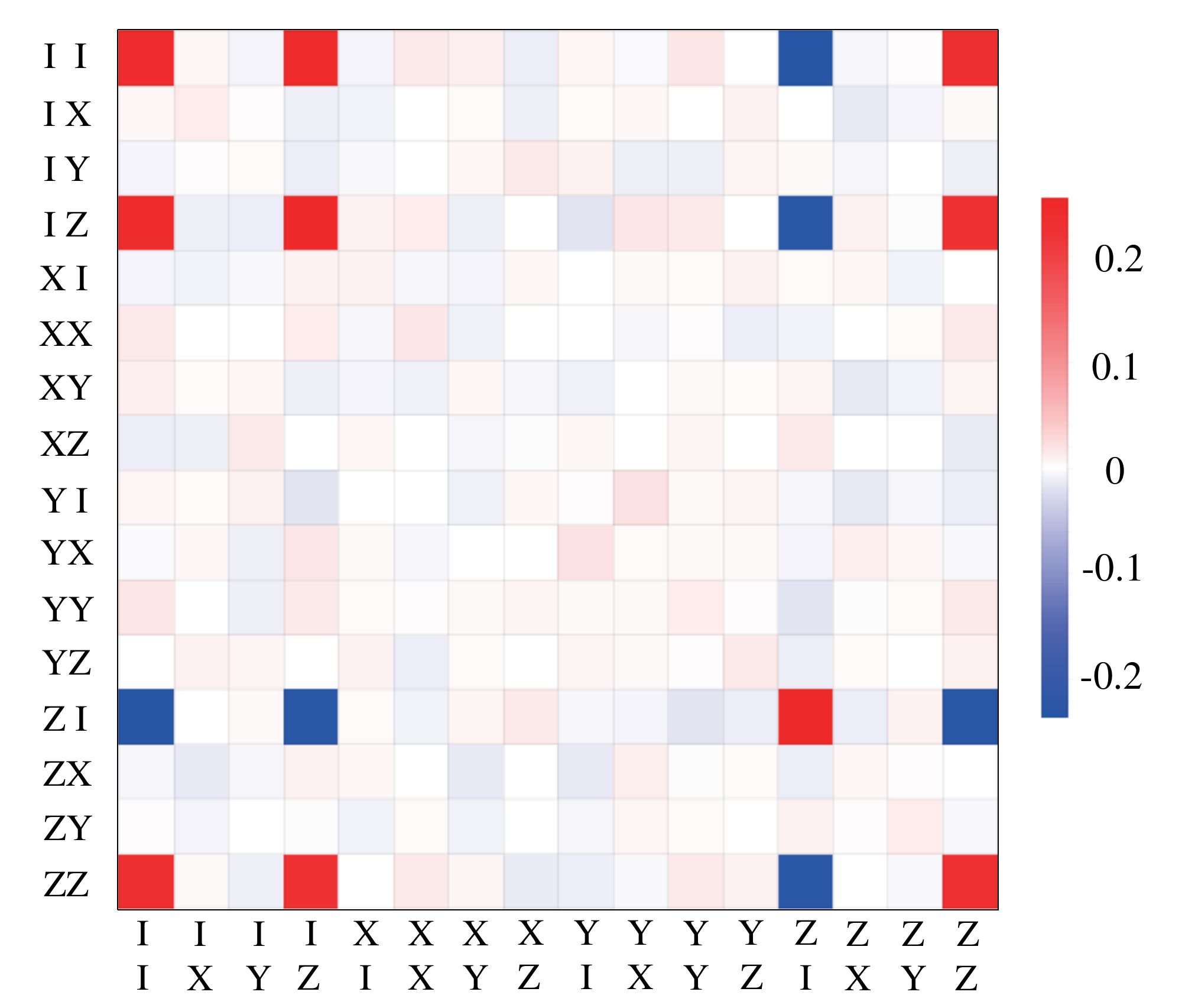}
	\caption{Real parts of the $\chi$ matrix of the C-phase Gate in spin basis. Theoretical imaginary parts equal zero, absolute values of the measured imaginary parts are all smaller than $0.04$ and are not shown.}
	\label{Cphasechi}
\end{figure}

The QPT is further applied to characterize the performance of the implemented gate operations. For example, the $\chi$ matrices of Hadmard$_1$ and C-phase $=$ diag$(1,-1,1,1)$ (spin basis) obtained by QPT  are illustrated in fig. \ref{H1chi} and \ref{Cphasechi}. Here the mapping operator is already applied to obtain the $\chi$ matrices in spin basis. The fidelities are $97.5$\% and $98.5$\%, respectively.

Although the control pulses obtained by GRAPE are already made robust against pulse amplitude errors \cite{khaneja2005optimal}, the fidelities are still below ideal values. We believe the main limitation is due to  magnetic noises, which randomly shift the energy levels (especially the Zeeman sublevels $|1,4\rangle$). This causes pulse errors  both during gate operations and tomographic pulses, and also degrades the coherence among the four levels. Reducing magnetic noises via triggering to the ac-line  \cite{ruster2016long} increases the control performance, as indicated by our experiment: the state fidelities of Hadamard$_{1,2}$ with $300$ $\mu$s gate times are increased from around $85\%$ without line triggering, to around $99\%$ with line triggering. Higher fidelities could be envisioned with magnetic shielding.

\section{Application: Demonstration of the Grover's algorithm}
\begin{figure}
	\centering
	\includegraphics[width=0.48\textwidth]{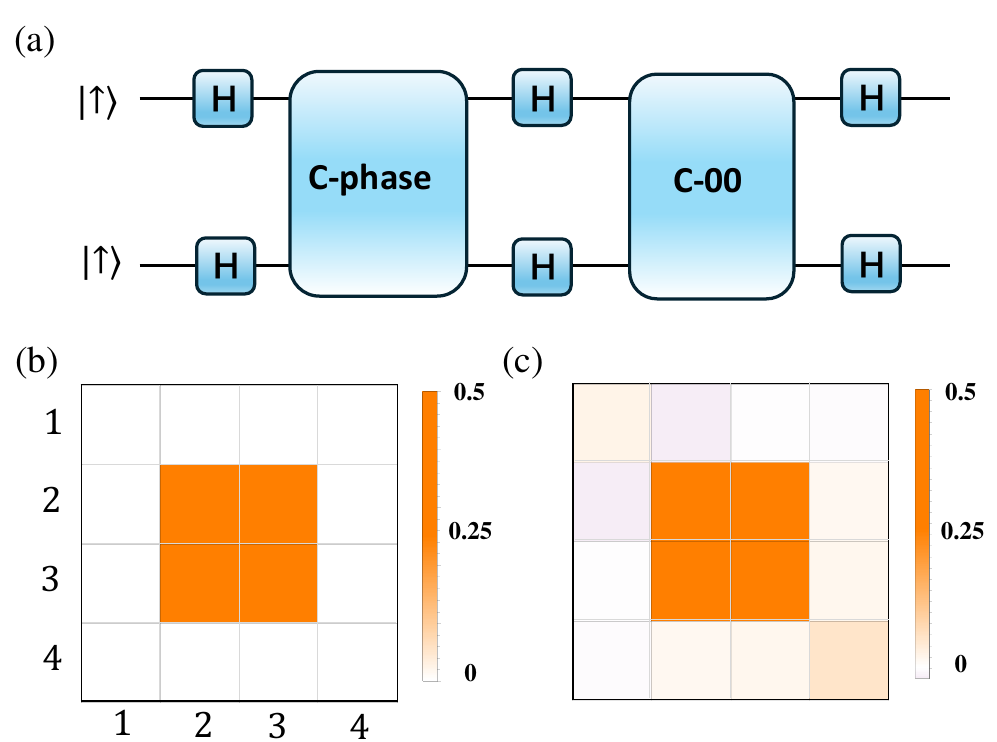}
	\caption{Implementation of the Grover's algorithm. (a) The quantum circuit. (b) Real parts of the theoretical density matrix of the final state in number basis. Imaginary parts equal zero. (c) Real part of the experimental density matrix of the final state. Absolute values of measured imaginary parts are all smaller than $0.06$ and are not shown. }
	\label{grover}
\end{figure}

Based on the universal $1$-ion-$2$-qubit processor, we could realize simple quantum algorithms, for example, the Grover's algorithm \cite{nielsen2002quantum}. As demonstrated in fig. \ref{grover}(a), the quantum circuit that searches for $\left|\uparrow \downarrow \right\rangle$ is implemented. The initial state $\left|\uparrow \uparrow\right\rangle$ is prepared by optical pumping to $|3\rangle$ then apply a $\pi$ pulse, $\textrm{C-00}=\textrm{diag}(1,-1,-1,-1)$ in spin basis.
 The theoretical and measured density matrix of the final state (in number basis) are shown in fig. \ref{grover}(b) and fig. \ref{grover}(c). The average successful rate is around  $99.0\%$.

\section{Scale up to $2n$-qubit universal quantum processors with $n$ ions}
Here we discuss the scalability of our scheme. That is, in an $n$-ion-$2n$-qubit processor, demonstrate how to realize arbitrary single-qubit gates and two-qubit entangling gates, not only within an ion, but also between two ions. Moreover, demonstrate how to readout the quantum information (state preparation is done by pumping all the ions to $|3\rangle$ and apply global $\pi$ pulses on $|3\rangle \leftrightarrow |1\rangle$). The fact that the two qubits are naturally existing electron and nuclear spins with hyperfine coupling allows novel and efficient scaling protocols. 
To simplify discussion, we define $I^{i}_{j,k}$ as the spin momentum operator in $k$ direction for qubit $j$ in ion $i$ [i.e., qubit-1(2)  means the nuclear (electron) spin qubit in this ion], define $I^{i}_{j\leftrightarrow l,k}$ as the spin momentum operator in $k$ direction for transition $|j\rangle \leftrightarrow |l\rangle$ in ion $i$, define $R^{i}_{j\leftrightarrow l,k}(\theta)$ ($R^{i}_{j,k}(\theta)$) as the rotation operator in ion $i$ for $|j\rangle \leftrightarrow |l\rangle$ transition (qubit $j$) with an angle $\theta$ along $k$.

\subsection{Laser-less approach}
Universal trapped-ion quantum processor based on static or time-varying magnetic field gradient has been proposed and realized \cite{mintert2001ion,ospelkaus2008trapped,sutherland2019versatile,ospelkaus2011microwave,srinivas2021high}. Our scheme based on electron/nuclear spin-qubits posses unique advantages when combining with this approach. Here we use one time-varying gradient field near a normal mode frequency as an example \cite{ospelkaus2008trapped}.

As demonstrated in \cite{warring2013individual}, single-ion control ($|2\rangle \leftrightarrow |3\rangle$ subspace) could be accomplished by selective positioning of the ions in a spatially varying microwave magnetic field. Similarly, as transitions $|3\rangle \leftrightarrow |1,4\rangle$ are driven by $B_{\perp}$ (this is clear form \eqref{ahcrwa} in Appendix A), the component of corresponding near-resonant microwave fields perpendicular to $B_0$, single-ion control in $|3\rangle \leftrightarrow |1,4\rangle$ subspace could be realized by placing the ions to be controlled in configurations $B_{\perp}\neq0$, and the remaining ions in $B_{\perp}=0$. 

Without loss of generality, to realize entangling operations between  qubit $2$ in ion $p$ and qubit $2$ in ion $w$,  
assume the ions are aligned along the $y$ axis, the trap axis orthogonal to $y$ are along $x$ and $z$ directions, and the quantization field is along $z$. Also denote $q^n_z$ the displacement along $z$ of ion $n$ relative to its equilibrium position ($q^n_z=0$). $q^n_z$ could be expanded as a summation over normal modes $q^n_z=\sum_{j}b^n_{j,z}\epsilon_j(a_j+a^{\dagger}_j)$, where $a^{\dagger}_j$, $a_j$ are normal-mode creation and annihilation operators, $b^n_{j,z}$ are dimensionless coefficients, $\epsilon_j=\sqrt{\hbar/(2m\omega_j)} \sim 1$ nm and $\omega_j$ are mode frequencies \cite{james1997quantum}. At positions with $z$ coordinates $q_z$, there is an oscillating magnetic field gradient $\vec{B}= q_zB'_z\cos(\omega t+\phi)\vec{e}_z$ that serves to couple the internal states and motion, $\omega=\omega_j-\delta$ close to a normal mode frequency. Only the ions $p,w$ are affected, this is done by selective positioning of the ions \cite{warring2013individual}. 
The total Hamiltonian in the lab frame reads $H_l=H^{p}_0+H^{w}_0+\sum_{j}\omega_ja^{\dagger}_ja_j+H_g$, where $H^{p}_0$ and $H^{w}_0$ are single-ion-two-qubit Hamiltonian discussed in \eqref{h00}, and the interaction with the gradient field is 
\begin{equation}
\begin{aligned}
H_g&=-\sum_{n=p,w} q^{n}_zB'_z\cos(\omega t+\phi)(\gamma_1I^{n}_{1,z}+\gamma_2I^{n}_{2,z}).
\end{aligned}
\end{equation}
Transforming to the rotating frame with respect to $H_r=H^{p}_0+H^{w}_0+\sum_{j}\omega_ja^{\dagger}_ja_j$, neglect fast oscillating terms, as well as terms $\sim B_0\gamma_{1,2}/A,B_0\gamma_{1,2}/A^2$ ($|B_0\gamma_{1,2}|\ll A$), the resulting Hamiltonian is
\begin{equation}\label{ionph}
	\begin{aligned}
	H_{a}&=-\sum_{n,l}\Omega^n_{j,z}I^{n}_{l,z}[e^{-i(\delta t-\phi)}a_j+e^{i(\delta t-\phi)}a^{\dagger}_j],
	\end{aligned}
\end{equation}
where 
\begin{equation}
\Omega^n_{j,z}=b^{n}_{j,z}B'_z\frac{\gamma_1+\gamma_2}{4}\epsilon_j.
\end{equation}
  Set the evolution time  $\tau=2k_1\pi/\delta$ ($k_{1}$ is an integer) to disentangle the internal states and motion \cite{sorensen2000entanglement}, the resulting evolution is a global entangling operation
   
\begin{equation}\label{uzz}
U_{zz}(\tau)=\textrm{exp}[\frac{2k_1\pi i}{\delta^2}(\sum_{n,l}\Omega^n_{j,z}I^{n}_{l,z})^2].
\end{equation}
With single-ion addressing available, the desired operation (up to a global phase factor $\theta_1=2k_1\pi /\delta^2\sum_{n}(\Omega^{n}_{j,z})^2$) could be realized by the following sequence
\begin{equation}\label{bevo}
\begin{aligned}
U_{2,2,zz}(\tau)=&\textrm{exp}[\frac{2k_1\pi i}{\delta^2}(8\Omega^p_{j,z}\Omega^w_{j,z}I^{p}_{2,z}I^{w}_{2,z})]\exp(i\theta_1)\\
=&U_{zz}(\tau)R^{p}_{1,y}(\pi)R^{w}_{1,y}(\pi)U_{zz}(\tau)R^{w}_{1,y}(-\pi)\\
&U_{zz}(\tau)R^{p}_{1,y}(\pi)R^{w}_{1,y}(\pi)U_{zz}(\tau)R^{w}_{1,y}(-\pi).
\end{aligned}
\end{equation}
Entangling operation between other combinations of qubits could be achieved by choosing proper single ion operations in \eqref{bevo}. Entangling operations with arbitrary spin operators, e.g., $U_{2,2,xx}$,
could be achieved by the sequence: $U_{2,2,xx}=R^{p}_{2,y}(\pi/2)R^{w}_{2,y}(\pi/2)U_{2,2,zz}R^{p}_{2,y}(-\pi/2)R^{w}_{2,y}(-\pi/2)$. As the 2-ion-4-qubit system, with the available control fields discussed above, is complete controllable, one could also achieve arbitrary operations via numerical optimal control.

$U_{2,2,zz}$ could also be achieved with only global control fields on both ions, provided that the quantization magnetic field is large ($|B_0\gamma_{1,2}|\gg A$). This is because given $H_l$, one could now transform to a rotating frame with respect to $\tilde{H}_r=H^{p}_z+H^{w}_z+A(I^{p}_{1,z}I^{p}_{2,z}+I^{w}_{1,z}I^{w}_{2,z})+\sum_{j}\omega_ja^{\dagger}_ja_j$, where $H^{n}_z=B_0(\gamma_{1}I^{n}_{1,z}+\gamma_{2}I^{n}_{2,z})$. After RWA 
\begin{equation}
\begin{aligned}
H_{a}&=-\sum_{n,l}\tilde{\Omega}^n_{l,j,z}I^{n}_{l,z}[e^{-i(\delta t-\phi)}a_j+e^{i(\delta t-\phi)}a^{\dagger}_j]\\
&\approx -\sum_{n}\tilde{\Omega}^n_{2,j,z}I^{n}_{2,z}[e^{-i(\delta t-\phi)}a_j+e^{i(\delta t-\phi)}a^{\dagger}_j],
\end{aligned}
\end{equation}
where 
\begin{equation}
\tilde{\Omega}^n_{l,j,z}=b^{n}_{j,z}B'_z\gamma_l\epsilon_j,
\end{equation}
and the approximation holds because $|\gamma_2| \gg |\gamma_1|$.  This is advantageous as it allows direct selective-entangling operation on qubit-2, with only global controls on both ions. Setting the evolution time  $\tau=2k_1\pi/(\delta)$ one obtains the desired entangling operation.
\subsection{Laser-based approach
}\label{lg}
Using single-ion addressing lasers with adjustable polarization, arbitrary single-ion-two-qubit operations could be achieved in an $n$-ion processor.  Arbitrary two-qubit entangling operation between qubits in different ions could be realized by combination of conventional Mølmer–Sørensen (MS) interaction with single-ion-two-qubit control pulses. This is  equivalent to the two-ion qudit entangling operation \cite{ringbauer2022universal,hrmo2023native} plus similar transformation. For example, the entangling operation $U_{1,1,xx}(\tau)=\textrm{exp}[-4i(I^{p}_{1,x}I^{w}_{1,x})\tau]$ could be realized by the following sequence:
\begin{equation}\label{msl}
	\begin{aligned}
U_{1,1,xx}(\tau)=&R^{\dagger}_5R_4\tilde{U}_{xx}(\tau)R^{\dagger}_4R_3\tilde{U}_{xx}(\tau)R^{\dagger}_3\\
&R_2\tilde{U}_{xx}(\tau)R^{\dagger}_2R_1\tilde{U}_{xx}(\tau)R^{\dagger}_1R_5,
\end{aligned}
\end{equation}
where
\begin{equation} 
\tilde{U}_{xx}(\tau)=\textrm{exp}[-i(I^{p}_{2\leftrightarrow 3,x}I^{w}_{2\leftrightarrow 3,x})\tau]
\end{equation}
is the conventional MS interaction between two clock qubits ($|2\rangle \leftrightarrow |3\rangle$ subspace of each ion), and
\begin{equation}\label{laserr}
\begin{aligned}
&R_1=R^{p}_{1\leftrightarrow 4,y}(\pi/2)R^{p}_{1\leftrightarrow 2,y}(\pi)R^{w}_{1\leftrightarrow 4,y}(\pi/2)R^{w}_{3\leftrightarrow 4,y}(\pi),\\
&R_2=R^{p}_{1\leftrightarrow 4,y}(\pi/2)R^{p}_{3\leftrightarrow 4,y}(\pi)R^{w}_{1\leftrightarrow 4,y}(\pi/2)R^{w}_{1\leftrightarrow 2,y}(\pi),\\
&R_3=R^{p}_{1\leftrightarrow 4,y}(\pi/2)R^{p}_{1\leftrightarrow 2,y}(\pi)R^{w}_{1\leftrightarrow 4,y}(\pi/2)R^{w}_{1\leftrightarrow 2,y}(\pi),\\
&R_4=R^{p}_{1\leftrightarrow 4,y}(\pi/2)R^{p}_{3\leftrightarrow 4,y}(\pi)R^{w}_{1\leftrightarrow 4,y}(\pi/2)R^{w}_{3\leftrightarrow 4,y}(\pi),\\
&R_5=R^{p}_{2\leftrightarrow 3,y}(\theta')R^{w}_{2\leftrightarrow 3,y}(\theta'),
\end{aligned}
\end{equation}
are single-ion operations, where
\begin{equation}
\theta'\approx 0=2\textrm{arctan}[\frac{\sqrt{2}\cos(\frac{\theta_0}{2})+\sqrt{2}\sin(\frac{\theta_0}{2})}{\sqrt{2}\cos(\frac{\theta_0}{2})-\sqrt{2}\sin(\frac{\theta_0}{2})}],
\end{equation}
and could be realized by our method with single-ion-addressing laser beams. Entangling gates with arbitrary spin operators  between arbitrary qubits could be similarly obtained. If the goal is to realize global entangling gates among, e.g., the four qubits in two ions, then only global control fields implemented simultaneously on both ions are required. It is thus promising to increase the number of spins when simulating Ising models, e.g., the $300$-spin system simulated in the recent work \cite{duan2024} could be readily upgraded to a $600$-spin system.

One could also apply hybrid scheme that uses both microwaves and lasers. For example, to apply entangling gates between qubits in ion $p,w$, one could first coherently transfer the states of these two ions to  metastable levels (e.g., $^2$F$_{7/2}$-subspace \cite{yang2022realizing,allcock2021omg,campbell2022polyqubit}), and apply the two-qubit gate via \eqref{msl}. While $\tilde{U}_{xx}(\tau)$ is still achieved by lasers, $R_{1-5}$ could now be achieved by microwaves: as the resonance frequencies in $^2$F$_{7/2}$ and $^2$S$_{1/2}$ subspaces are different, these operations will not affect other ions. At last, one transfer the states back to $^2$S$_{1/2}$ subspace. These transfer operations only require $\pi$ rotations and are relatively easy to implement.


\subsection{Readout}\label{readout}
If only a small number of ions is of interest, one could apply the method in sec. \ref{onetwo} and obtain the full density matrix through quantum state tomography, as long as the efficiency is not a concern. In general, in an $n$-ion-$2n$-qubit processor, the diagonal elements of the density matrix could be obtained by the same method in qudit processors \cite{low2020practical}: coherently transfer the states to a metastable manifold (e.g., $^2$D$_{5/2}$ or $^2$F$_{7/2}$-subspace \cite{yang2022realizing,allcock2021omg,campbell2022polyqubit}), and sequentially transfer them back to the ground state manifold and readout. These operations are applied simultaneously to all ions. 
In some cases, individual qubit measurement within an ion ($p$) is required, e.g., to implement mid-circuit error correction through measuring spectator qubits \cite{singh2023mid}.
One must measure the corresponding qubit without destroying the other one. This could be achieved by introducing an ancilla ion, applying motion-sensitive laser fields to ion $p$, and readout the qubit through measuring the phonon state of the ancilla ion \cite{campbell2022polyqubit}. Alternatively, one could apply a two-qubit SWAP gate that transfers the qubit state to the ancilla ion, and then measure the ancilla ion. Note interestingly that different spin states generate different motional states in a gradient magnetic field, moreover, this effect is significantly different for electron-qubit states and nuclear-qubit states ($|\gamma_e| \gg |\gamma_n|$). This provides the possibility to design more efficient readout protocols that utilize not only the ``brightness" of an ion, but also its spatial position to distinguish different qubit states within an ion.

\section{Conclusion}
In conclusion, we construct a scalable $n$-ion-$2n$-qubit quantum processor, where the four energy levels of a single ion is utilized to encode two qubits. We experimentally realize a  $1$-ion-$2$-qubit universal processor, where the valence electron spin and nuclear spin of a single $^{171}$Yb$^{+}$ ion constitute the two-qubit system. Process fidelities of single-qubit and two-qubit gates are around $98\%$. The Grover's algorithm is further implemented with a successful rate exceeding $99\%$. Higher fidelities could be achieved by more sophisticated control sequences and magnetic shielding (which also brings benefits to coherence time). The scaling-up scheme is straightforward, and we give explicit protocols based on both laser-less and laser-based frameworks. We also discuss the readout process, which could be achieved using well-known techniques. We further discover that under suitable relative-strength between the hyperfine interaction and interaction with quantization magnetic-field, the large discrepancy between $\gamma_e$ and $\gamma_n$ allows more efficient two-qubit entangling gates between different ions, and may be utilized to design novel readout schemes.

Our work paves the way to achieving more information processing capability ($2^{2n}$ Hilbert-space) using $n$ ions, comparable to the qudit $(d=4)$-based scheme. Compared to qudits, the two-qubit scheme could use standard, binary quantum algorithms without modification, including qubit quantum error correction \cite{campbell2022polyqubit}. Besides, the replacement of some inter-atomic gates  by intra-atomic gates in the two-qubit scheme could increase the fidelity of some quantum circuits. On the other hand, the similarities between these two schemes allow easy adaptation of control protocols.  The two qubits could also be constructed by any convenient four levels, e.g., the four clock states in $^2$S$_{1/2}$ and $^2$F$_{7/2}$. The scheme with electron/nuclear spin has clear two-qubit entanglement structure, and provides potential efficiency and novelty in certain tasks such as simulating lattice gauge theories \cite{banerjee2013atomic} and preparing thermofield double states \cite{cottrell2019build,zhu2020generation}. 

\section*{Acknowledgments}
This work is supported by the Key-Area Research and Development Program of Guangdong Province under Grant
No.2019B030330001, the National Natural Science Foundation of China under Grant No.11774436, No.11974434
and No.12074439, Natural Science Foundation of Guangdong Province under Grant 2020A1515011159, Science and Technology Program of Guangzhou, China
202102080380, the Shenzhen Science and Technology Program under Grant No.2021Szvup172 and
JCYJ20220818102003006. Le Luo acknowledges the support from Guangdong Province Youth Talent Program
under Grant No.2017GC010656. Ji Bian acknowledges
the support from the China Postdoctoral Science Foundation under Grant No.2021M703768.

\section*{Appendix A: Universal quantum control in the 1-ion-2-qubit processor }\label{auni}

Here we derive $H_c$ in \eqref{hcrwa} under microwave control fields. Applying a general near-resonant microwave magnetic field $\sum_{k}\mathbf{B}^{k}(t)\cos[\omega_k (t)t+\phi_{k}(t)]$ to the ion, we get the Hamiltonian in the lab frame 
\begin{equation}
	H'_l=H_0-\sum_k\sum_{j=x,y,z}B^{k}_{j}\cos(\omega_k t+\phi_k)(\gamma_1I_{1,j}+\gamma_2I_{2,j}),
\end{equation}
where $B^{k}_{x,y,z}(t)\cos(\omega_k t+\phi_{k})$ are components along $x,y,z$ axis.
Assume the following parameter regime which describes the actual experiment:  $|\omega_{1,2,3}-(E_{1,2,4}-E_3)| \ll E_1-E_2$ (and $E_2-E_4$), $|\gamma_{1,2}B^{k}_{x,y,z}| \ll E_1-E_2$ (and $E_2-E_4$),  $\omega_{1,2,3} \gg$ any combination of $|\gamma_{1,2}B^{k}_{x,y,z}|$, and $B^{k}_{x,y,z}(t)$ varies slowly compared to $\omega_{1,2,3}$. Set $\omega_1=E_1-E_3-\tilde{\delta}_1(t)$, $\omega_2=E_2-E_3-\tilde{\delta}_2(t)$, $\omega_3=E_4-E_3-\tilde{\delta}_4(t)$, where $\tilde{\delta}_{1,2,4}(t)=1/t\int_{0}^{t}\delta_{1,2,4}(s)ds$. Then transforming to the rotating frame $r'$ with respect to $H'_{r'}(t)=(\delta_1(t)-E_1)|1\rangle \langle 1 |+(\delta_2(t)-E_2)|2\rangle \langle 2 |-E_3|3\rangle \langle 3 |+(\delta_4(t)-E_4)|4\rangle \langle 4 |$ and apply the RWA, we get
\begin{equation}\label{ahcrwa}
	\begin{aligned}
		H_c&=R_{r'}H'_lR^{\dagger}_{r'}+i\dot{R}_{r'}R^{\dagger}_{r'}\\
		&\approx \frac{\delta_1(t)}{2}|1\rangle \langle 1 |+\frac{\delta_2(t)}{2}|2\rangle \langle 2 |+\frac{\delta_4(t)}{2}|4\rangle \langle 4 |\\
		&+\frac{1}{2}(B^1_x+iB^1_y)(\gamma_1\cos\frac{\theta_0}{2}+\gamma_2\sin\frac{\theta_0}{2})e^{i\phi_1}|3\rangle \langle 1 |\\
		&-\frac{1}{2}B^2_z\sin\frac{\theta_0}{2}\cos\frac{\theta_0}{2}(\gamma_2-\gamma_1)e^{i\phi_2}|3\rangle \langle 2 |\\
		&+\frac{1}{2}(B^3_x-iB^3_y)(\gamma_1\sin\frac{\theta_0}{2}+\gamma_2\cos\frac{\theta_0}{2})e^{i\phi_3}|3\rangle \langle 4 |\\
		&+\textrm{H.c.},
	\end{aligned}
\end{equation}
where
$R_{r'}(t)=\textrm{exp}(-i\int_{0}^{t}H'_{r'}(s)ds)$.
Compared to \eqref{hcrwa}, we could identify $c_{31,32,34}(t)$, $\delta_{1,2,4}(t)$, and implement them through modulating the amplitudes, frequencies and phases of the microwave field. It is also clear that the component of the microwave field parallel (perpendicular) to the quantization axis drives the $|3\rangle \leftrightarrow |2\rangle$ ($|3\rangle \leftrightarrow |1,4\rangle$) transition.

$R_{r'}$ is itself an entangling operation on electron spin and nuclear spin, so the degree of entanglement (e.g., the entanglement of formation \cite{wootters1998entanglement}) between these two spins in $r'$ is different from that in the lab frame $l$. Nevertheless, the entanglement in $r'$ is physically well defined, as one could straightforwardly transfer from $r'$ to $l$ and obtain the ``true degree of entanglement''. If necessary, it is also straightforward to implement an entangling operation $U_l$ in $l$: one could just search for pulses that implement $R_{r'}U_l$ in $r'$ by GRAPE. 

\section*{Appendix B: Characterization of the two-qubit quantum processor}\label{aexp}

\begin{figure}[htb] 

	\includegraphics[scale=0.27]{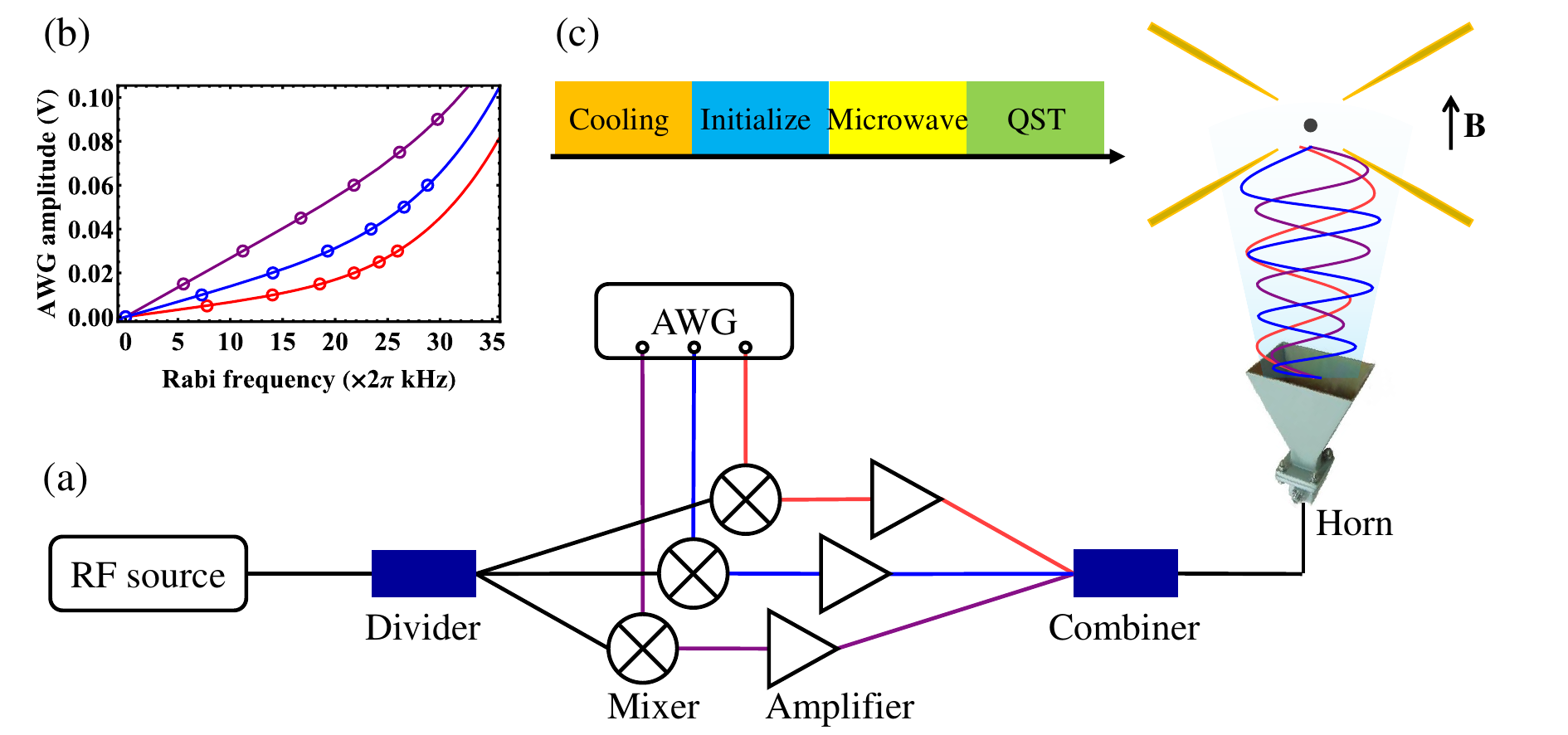}
	\caption{Experimental setup of the two-qubit quantum processor, with purple representing $|3\rangle \leftrightarrow |2\rangle$, blue representing $|3\rangle \leftrightarrow |1\rangle$, and red representing $|3\rangle \leftrightarrow |4\rangle$. (a) Microwave circuits utilized in the experiment. (b) Calibration curves of the AWG amplitudes versus Rabi frequencies. (c) The timing sequence of the experiment.}
	\label{microwave}
\end{figure}

The ion is confined near the center of the trap, which consists of four blade electrodes. The ion is trapped through the application of radio frequency (RF) and direct current (DC) voltages. The quantization magnetic field ($B_0 \approx 6$ Gs) is provided by the current in Helmholtz coils and is oriented along the Z-axis. It is further stabilized by a PI controller, ensuring stable Zeeman levels. The microwave signals ($\approx 2\pi \times 12.6$ GHz) utilized to drive the three transitions $|3\rangle \leftrightarrow |1,2,4\rangle$ are generated by mixing a standard RF source with an arbitrary waveform generator (AWG) (see fig. \ref{microwave}(a)). Notably, the signal from the RF source is divided into three paths by a power divider and subsequently mixed with the three paths of the AWG. Each of the three resulting microwaves is then amplified individually before being combined into a single path. This generates large Rabi frequencies while avoiding the creation of high-order harmonics. Finally, the combined signal is directed into the trap via the microwave horn. The calibration of the microwave amplitudes versus Rabi frequencies for the transitions $|3\rangle \leftrightarrow |1,2,4\rangle$ is illustrated in fig. \ref{microwave}(b). The coherence time $T^{*}_2$ of the relevant transitions measured by Ramsey sequences are listed in table \ref{tab:CT1}. Using triggering to AC-line techniques \cite{ruster2016long}, the coherence time could be significantly increased, as listed in table \ref{tab:CT1}.  Fig.\ref{microwave} (c) depicts our experimental sequence. Initially, the ion is cooled by Doppler-cooling beams. Subsequently, the state is initialized to $|3\rangle$ by optical pumping. Next, the system is manipulated by the three-path microwave. Finally, QST is performed to obtain the full density matrix of the final state.

\begin{table}[h]
	\caption{Coherence time without (first row) and with (second row) line triggering.}
	\begin{tabular}{ccccc}
		\hline \hline 
		  & $|2\rangle \leftrightarrow |3\rangle$ & $|1\rangle \leftrightarrow |3\rangle$ & $|4\rangle \leftrightarrow |3\rangle$  & $|1\rangle \leftrightarrow|4\rangle$  \\ \hline	                                              
		 & $20$ ms & $500$ $\mu$s & $500$ $\mu$s & $200$ $\mu$s  \\
		& $20$ ms   & $7$ ms  & $7$ ms & $3$ ms   \\
		 \hline \hline
	\end{tabular}\label{tab:CT1}
\end{table}


\begin{thebibliography}{10}
	
	\bibitem{bruzewicz2019trapped}
	Colin~D Bruzewicz, John Chiaverini, Robert McConnell, and Jeremy~M Sage.
	\newblock Trapped-ion quantum computing: Progress and challenges.
	\newblock {\em Applied Physics Reviews}, 6(2):021314, 2019.
	
	\bibitem{monroe2021programmable}
	Christopher Monroe, Wes~C Campbell, L-M Duan, Z-X Gong, Alexey~V Gorshkov,
	Paul~W Hess, Rajibul Islam, Kihwan Kim, Norbert~M Linke, Guido Pagano, et~al.
	\newblock Programmable quantum simulations of spin systems with trapped ions.
	\newblock {\em Reviews of Modern Physics}, 93(2):025001, 2021.
	
	\bibitem{saffman2010quantum}
	Mark Saffman, Thad~G Walker, and Klaus M{\o}lmer.
	\newblock Quantum information with rydberg atoms.
	\newblock {\em Reviews of modern physics}, 82(3):2313, 2010.
	
	\bibitem{cai2023entangling}
	Zhengyang Cai, Chun-Yang Luan, Lingfeng Ou, Hengchao Tu, Zihan Yin, Jing-Ning
	Zhang, and Kihwan Kim.
	\newblock Entangling gates for trapped-ion quantum computation and quantum
	simulation.
	\newblock {\em Journal of the Korean Physical Society}, 82(9):882--900, 2023.
	
	\bibitem{devoret2013superconducting}
	Michel~H Devoret and Robert~J Schoelkopf.
	\newblock Superconducting circuits for quantum information: an outlook.
	\newblock {\em Science}, 339(6124):1169--1174, 2013.
	
	\bibitem{clarke2008superconducting}
	John Clarke and Frank~K Wilhelm.
	\newblock Superconducting quantum bits.
	\newblock {\em Nature}, 453(7198):1031--1042, 2008.
	
	\bibitem{you2005superconducting}
	JQ~You and Franco Nori.
	\newblock Superconducting circuits and quantum information.
	\newblock {\em Physics today}, 58(11):42--47, 2005.
	
	\bibitem{preskill2018quantum}
	John Preskill.
	\newblock Quantum computing in the nisq era and beyond.
	\newblock {\em Quantum}, 2:79, 2018.
	
	\bibitem{arute2019quantum}
	Frank Arute, Kunal Arya, Ryan Babbush, Dave Bacon, Joseph~C Bardin, Rami
	Barends, Rupak Biswas, Sergio Boixo, Fernando~GSL Brandao, David~A Buell,
	et~al.
	\newblock Quantum supremacy using a programmable superconducting processor.
	\newblock {\em Nature}, 574(7779):505--510, 2019.
	
	\bibitem{zhong2020quantum}
	Han-Sen Zhong, Hui Wang, Yu-Hao Deng, Ming-Cheng Chen, Li-Chao Peng, Yi-Han
	Luo, Jian Qin, Dian Wu, Xing Ding, Yi~Hu, et~al.
	\newblock Quantum computational advantage using photons.
	\newblock {\em Science}, 370(6523):1460--1463, 2020.
	
	\bibitem{wang2020qudits}
	Yuchen Wang, Zixuan Hu, Barry~C Sanders, and Sabre Kais.
	\newblock Qudits and high-dimensional quantum computing.
	\newblock {\em Frontiers in Physics}, 8:589504, 2020.
	
	\bibitem{ringbauer2022universal}
	Martin Ringbauer, Michael Meth, Lukas Postler, Roman Stricker, Rainer Blatt,
	Philipp Schindler, and Thomas Monz.
	\newblock A universal qudit quantum processor with trapped ions.
	\newblock {\em Nature Physics}, 18(9):1053--1057, 2022.
	
	\bibitem{hrmo2023native}
	Pavel Hrmo, Benjamin Wilhelm, Lukas Gerster, Martin~W van Mourik, Marcus Huber,
	Rainer Blatt, Philipp Schindler, Thomas Monz, and Martin Ringbauer.
	\newblock Native qudit entanglement in a trapped ion quantum processor.
	\newblock {\em Nature Communications}, 14(1):2242, 2023.
	
	\bibitem{liu2023performing}
	Pei Liu, Ruixia Wang, Jing-Ning Zhang, Yingshan Zhang, Xiaoxia Cai, Huikai Xu,
	Zhiyuan Li, Jiaxiu Han, Xuegang Li, Guangming Xue, et~al.
	\newblock Performing su (d) operations and rudimentary algorithms in a
	superconducting transmon qudit for d= 3 and d= 4.
	\newblock {\em Physical Review X}, 13(2):021028, 2023.
	
	\bibitem{macdonell2021analog}
	Ryan~J MacDonell, Claire~E Dickerson, Clare~JT Birch, Alok Kumar, Claire~L
	Edmunds, Michael~J Biercuk, Cornelius Hempel, and Ivan Kassal.
	\newblock Analog quantum simulation of chemical dynamics.
	\newblock {\em Chemical Science}, 12(28):9794--9805, 2021.
	
	\bibitem{rico2018so}
	Enrique Rico, Marcello Dalmonte, Peter Zoller, Debasish Banerjee, Michael
	B{\"o}gli, Pascal Stebler, and U-J Wiese.
	\newblock So (3)“nuclear physics” with ultracold gases.
	\newblock {\em Annals of physics}, 393:466--483, 2018.
	
	\bibitem{campbell2022polyqubit}
	Wesley~C Campbell and Eric~R Hudson.
	\newblock Polyqubit quantum processing.
	\newblock {\em arXiv preprint arXiv:2210.15484}, 2022.
	
	\bibitem{yang2022realizing}
	H-X Yang, J-Y Ma, Y-K Wu, Ye~Wang, M-M Cao, W-X Guo, Y-Y Huang, Lu~Feng, Z-C
	Zhou, and L-M Duan.
	\newblock Realizing coherently convertible dual-type qubits with the same ion
	species.
	\newblock {\em Nature Physics}, 18(9):1058--1061, 2022.
	
	\bibitem{simm2023two}
	Alexander Simm, Shai Machnes, and Frank~K Wilhelm.
	\newblock Two qubits in one transmon--qec without ancilla hardware.
	\newblock {\em arXiv preprint arXiv:2302.14707}, 2023.
	
	\bibitem{prawer2008diamond}
	Steven Prawer and Andrew~D Greentree.
	\newblock Diamond for quantum computing.
	\newblock {\em Science}, 320(5883):1601--1602, 2008.
	
	\bibitem{wu2019programmable}
	Yang Wu, Ya~Wang, Xi~Qin, Xing Rong, and Jiangfeng Du.
	\newblock A programmable two-qubit solid-state quantum processor under ambient
	conditions.
	\newblock {\em npj Quantum Information}, 5(1):9, 2019.
	
	\bibitem{castelletto2020silicon}
	Stefania Castelletto and Alberto Boretti.
	\newblock Silicon carbide color centers for quantum applications.
	\newblock {\em Journal of Physics: Photonics}, 2(2):022001, 2020.
	
	\bibitem{debnath2016demonstration}
	Shantanu Debnath, Norbert~M Linke, Caroline Figgatt, Kevin~A Landsman, Kevin
	Wright, and Christopher Monroe.
	\newblock Demonstration of a small programmable quantum computer with atomic
	qubits.
	\newblock {\em Nature}, 536(7614):63--66, 2016.
	
	\bibitem{mintert2001ion}
	Florian Mintert and Christof Wunderlich.
	\newblock Ion-trap quantum logic using long-wavelength radiation.
	\newblock {\em Physical Review Letters}, 87(25):257904, 2001.
	
	\bibitem{ospelkaus2008trapped}
	Christian Ospelkaus, Christopher~E Langer, Jason~M Amini, Kenton~R Brown,
	Dietrich Leibfried, and David~J Wineland.
	\newblock Trapped-ion quantum logic gates based on oscillating magnetic fields.
	\newblock {\em Physical review letters}, 101(9):090502, 2008.
	
	\bibitem{sutherland2019versatile}
	RT~Sutherland, Raghavendra Srinivas, Shaun~C Burd, Dietrich Leibfried, Andrew~C
	Wilson, David~J Wineland, DTC Allcock, DH~Slichter, and SB~Libby.
	\newblock Versatile laser-free trapped-ion entangling gates.
	\newblock {\em New Journal of Physics}, 21(3):033033, 2019.
	
	\bibitem{ospelkaus2011microwave}
	C~Ospelkaus, U~Warring, Y~Colombe, KR~Brown, JM~Amini, D~Leibfried, and David~J
	Wineland.
	\newblock Microwave quantum logic gates for trapped ions.
	\newblock {\em Nature}, 476(7359):181--184, 2011.
	
	\bibitem{srinivas2021high}
	Raghavendra Srinivas, SC~Burd, HM~Knaack, RT~Sutherland, Alex Kwiatkowski,
	Scott Glancy, Emanuel Knill, DJ~Wineland, Dietrich Leibfried, Andrew~C
	Wilson, et~al.
	\newblock High-fidelity laser-free universal control of trapped ion qubits.
	\newblock {\em Nature}, 597(7875):209--213, 2021.
	
	\bibitem{wang2022fast}
	Kaizhao Wang, Jing-Fan Yu, Pengfei Wang, Chunyang Luan, Jing-Ning Zhang, and
	Kihwan Kim.
	\newblock Fast multi-qubit global-entangling gates without individual
	addressing of trapped ions.
	\newblock {\em Quantum Science and Technology}, 7(4):044005, 2022.
	
	\bibitem{wong2017demonstration}
	Jamie~David Wong-Campos, Steven~A Moses, Kale~Gifford Johnson, and Christopher
	Monroe.
	\newblock Demonstration of two-atom entanglement with ultrafast optical pulses.
	\newblock {\em Physical Review Letters}, 119(23):230501, 2017.
	
	\bibitem{schafer2018fast}
	VM~Sch{\"a}fer, CJ~Ballance, K~Thirumalai, LJ~Stephenson, TG~Ballance,
	AM~Steane, and DM~Lucas.
	\newblock Fast quantum logic gates with trapped-ion qubits.
	\newblock {\em Nature}, 555(7694):75--78, 2018.
	
	\bibitem{de2016parallel}
	Ludwig~E de~Clercq, Hsiang-Yu Lo, Matteo Marinelli, David Nadlinger, Robin
	Oswald, Vlad Negnevitsky, Daniel Kienzler, Ben Keitch, and Jonathan~P Home.
	\newblock Parallel transport quantum logic gates with trapped ions.
	\newblock {\em Physical Review Letters}, 116(8):080502, 2016.
	
	\bibitem{low2020practical}
	Pei~Jiang Low, Brendan~M White, Andrew~A Cox, Matthew~L Day, and Crystal Senko.
	\newblock Practical trapped-ion protocols for universal qudit-based quantum
	computing.
	\newblock {\em Physical Review Research}, 2(3):033128, 2020.
	
	\bibitem{baldwin2021high}
	CH~Baldwin, BJ~Bjork, M~Foss-Feig, JP~Gaebler, D~Hayes, MG~Kokish, C~Langer,
	JA~Sedlacek, D~Stack, and G~Vittorini.
	\newblock High-fidelity light-shift gate for clock-state qubits.
	\newblock {\em Physical Review A}, 103(1):012603, 2021.
	
	\bibitem{wang2011quantum}
	KL~Wang, Michael Johanning, Mang Feng, Florian Mintert, and Christof
	Wunderlich.
	\newblock Quantum gates using electronic and nuclear spins of yb+ in a magnetic
	field gradient.
	\newblock {\em The European Physical Journal D}, 63:157--164, 2011.
	
	\bibitem{feng2009nuclear}
	M~Feng, YY~Xu, F~Zhou, and D~Suter.
	\newblock Nuclear spin qubits in a trapped-ion quantum computer.
	\newblock {\em Physical Review A}, 79(5):052331, 2009.
	
	\bibitem{brownnutt2015ion}
	M~Brownnutt, M~Kumph, P~Rabl, and R~Blatt.
	\newblock Ion-trap measurements of electric-field noise near surfaces.
	\newblock {\em Reviews of modern Physics}, 87(4):1419, 2015.
	
	\bibitem{banerjee2013atomic}
	Debasish Banerjee, Michael B{\"o}gli, Marcello Dalmonte, Enrique Rico, Pascal
	Stebler, U-J Wiese, and Peter Zoller.
	\newblock Atomic quantum simulation of u (n) and su (n) non-abelian lattice
	gauge theories.
	\newblock {\em Physical review letters}, 110(12):125303, 2013.
	
	\bibitem{cottrell2019build}
	William Cottrell, Ben Freivogel, Diego~M Hofman, and Sagar~F Lokhande.
	\newblock How to build the thermofield double state.
	\newblock {\em Journal of High Energy Physics}, 2019(2):1--43, 2019.
	
	\bibitem{zhu2020generation}
	Daiwei Zhu, Sonika Johri, Norbert~M Linke, KA~Landsman, C~Huerta~Alderete,
	Nhunh~H Nguyen, AY~Matsuura, TH~Hsieh, and Christopher Monroe.
	\newblock Generation of thermofield double states and critical ground states
	with a quantum computer.
	\newblock {\em Proceedings of the National Academy of Sciences},
	117(41):25402--25406, 2020.
	
	\bibitem{serbyn2021quantum}
	Maksym Serbyn, Dmitry~A Abanin, and Zlatko Papi{\'c}.
	\newblock Quantum many-body scars and weak breaking of ergodicity.
	\newblock {\em Nature Physics}, 17(6):675--685, 2021.
	
	\bibitem{zhang2023many}
	Pengfei Zhang, Hang Dong, Yu~Gao, Liangtian Zhao, Jie Hao, Jean-Yves Desaules,
	Qiujiang Guo, Jiachen Chen, Jinfeng Deng, Bobo Liu, et~al.
	\newblock Many-body hilbert space scarring on a superconducting processor.
	\newblock {\em Nature Physics}, 19(1):120--125, 2023.
	
	\bibitem{foot2004atomic}
	Christopher~J Foot.
	\newblock {\em Atomic physics}, volume~7.
	\newblock OUP Oxford, 2004.
	
	\bibitem{yuan2022preserving}
	Xinxing Yuan, Yue Li, Mengxiang Zhang, Chang Liu, Mingdong Zhu, Xi~Qin,
	Nikolay~V Vitanov, Yiheng Lin, and Jiangfeng Du.
	\newblock Preserving multilevel quantum coherence by dynamical decoupling.
	\newblock {\em Physical Review A}, 106(2):022412, 2022.
	
	\bibitem{ruster2016long}
	Thomas Ruster, Christian~T Schmiegelow, Henning Kaufmann, Claudia Warschburger,
	Ferdinand Schmidt-Kaler, and Ulrich~G Poschinger.
	\newblock A long-lived zeeman trapped-ion qubit.
	\newblock {\em Applied Physics B}, 122(10):254, 2016.
	
	\bibitem{nielsen2002quantum}
	Michael~A Nielsen and Isaac Chuang.
	\newblock Quantum computation and quantum information, 2002.
	
	\bibitem{vandersypen2005nmr}
	Lieven~MK Vandersypen and Isaac~L Chuang.
	\newblock Nmr techniques for quantum control and computation.
	\newblock {\em Reviews of modern physics}, 76(4):1037, 2005.
	
	\bibitem{bian2017universal}
	Ji~Bian, Min Jiang, Jiangyu Cui, Xiaomei Liu, Botao Chen, Yunlan Ji, Bo~Zhang,
	John Blanchard, Xinhua Peng, and Jiangfeng Du.
	\newblock Universal quantum control in zero-field nuclear magnetic resonance.
	\newblock {\em Physical Review A}, 95(5):052342, 2017.
	
	\bibitem{jiang2018numerical}
	Min Jiang, Ji~Bian, Xiaomei Liu, Hengyan Wang, Yunlan Ji, Bo~Zhang, Xinhua
	Peng, and Jiangfeng Du.
	\newblock Numerical optimal control of spin systems at zero magnetic field.
	\newblock {\em Physical Review A}, 97(6):062118, 2018.
	
	\bibitem{jiang2018experimental}
	Min Jiang, Teng Wu, John~W Blanchard, Guanru Feng, Xinhua Peng, and Dmitry
	Budker.
	\newblock Experimental benchmarking of quantum control in zero-field nuclear
	magnetic resonance.
	\newblock {\em Science advances}, 4(6):eaar6327, 2018.
	
	\bibitem{albertini2002lie}
	Francesca Albertini and Domenico D'Alessandro.
	\newblock The lie algebra structure and controllability of spin systems.
	\newblock {\em Linear algebra and its applications}, 350(1-3):213--235, 2002.
	
	\bibitem{glaser2015training}
	Steffen~J Glaser, Ugo Boscain, Tommaso Calarco, Christiane~P Koch, Walter
	K{\"o}ckenberger, Ronnie Kosloff, Ilya Kuprov, Burkhard Luy, Sophie Schirmer,
	Thomas Schulte-Herbr{\"u}ggen, et~al.
	\newblock Training schr{\"o}dinger’s cat: Quantum optimal control: Strategic
	report on current status, visions and goals for research in europe.
	\newblock {\em The European Physical Journal D}, 69:1--24, 2015.
	
	\bibitem{lu2024experimental}
	Pengfei Lu, Xinxin Rao, Teng Liu, Yang Liu, Ji~Bian, Feng Zhu, and Le~Luo.
	\newblock Experimental demonstration of enhanced violations of leggett-garg
	inequalities in a pt-symmetric trapped-ion qubit.
	\newblock {\em Physical Review A}, 109(4):042205, 2024.
	
	\bibitem{bian2023quantum}
	Ji~Bian, Pengfei Lu, Teng Liu, Hao Wu, Xinxin Rao, Kunxu Wang, Qifeng Lao, Yang
	Liu, Feng Zhu, and Le~Luo.
	\newblock Quantum simulation of a general anti-pt-symmetric hamiltonian with a
	trapped ion qubit.
	\newblock {\em Fundamental Research}, 3(6):904--908, 2023.
	
	\bibitem{bian2023implementation}
	Ji~Bian, Teng Liu, Pengfei Lu, Qifeng Lao, Xinxin Rao, Feng Zhu, Yang Liu, and
	Le~Luo.
	\newblock Implementation of electromagnetic analogy to gravity mediated
	entanglement.
	\newblock {\em arXiv preprint arXiv:2304.06996}, 2023.
	
	\bibitem{lu2024realizing}
	Pengfei Lu, Teng Liu, Yang Liu, Xinxin Rao, Qifeng Lao, Hao Wu, Feng Zhu, and
	Le~Luo.
	\newblock Realizing quantum speed limit in open system with a-symmetric
	trapped-ion qubit.
	\newblock {\em New Journal of Physics}, 26(1):013043, 2024.
	
	\bibitem{zhao2022efficient}
	Fa~Zhao, Teng Liu, Ji~Bian, Peng-Fei Lu, Yang Liu, Feng Zhu, Xue-Ke Song, Dong
	Wang, Liu Ye, and Le~Luo.
	\newblock Efficient and robust chiral resolution based on non-adiabatic
	holonomic quantum computation.
	\newblock {\em arXiv preprint arXiv:2210.11740}, 2022.
	
	\bibitem{zhang2015time}
	Xiang Zhang, Yangchao Shen, Junhua Zhang, Jorge Casanova, Lucas Lamata, Enrique
	Solano, Man-Hong Yung, Jing-Ning Zhang, and Kihwan Kim.
	\newblock Time reversal and charge conjugation in an embedding quantum
	simulator.
	\newblock {\em Nature communications}, 6(1):7917, 2015.
	
	\bibitem{hu2018experimental}
	Chang-Kang Hu, Jin-Ming Cui, Alan~C Santos, Yun-Feng Huang, Marcelo~S Sarandy,
	Chuan-Feng Li, and Guang-Can Guo.
	\newblock Experimental implementation of generalized transitionless quantum
	driving.
	\newblock {\em Optics Letters}, 43(13):3136--3139, 2018.
	
	\bibitem{zhang2017observation}
	Jiehang Zhang, Paul~W Hess, A~Kyprianidis, Petra Becker, A~Lee, J~Smith,
	Gaetano Pagano, I-D Potirniche, Andrew~C Potter, Ashvin Vishwanath, et~al.
	\newblock Observation of a discrete time crystal.
	\newblock {\em Nature}, 543(7644):217--220, 2017.
	
	\bibitem{machnes2018tunable}
	Shai Machnes, Elie Ass{\'e}mat, David Tannor, and Frank~K Wilhelm.
	\newblock Tunable, flexible, and efficient optimization of control pulses for
	practical qubits.
	\newblock {\em Physical review letters}, 120(15):150401, 2018.
	
	\bibitem{doria2011optimal}
	Patrick Doria, Tommaso Calarco, and Simone Montangero.
	\newblock Optimal control technique for many-body quantum dynamics.
	\newblock {\em Physical review letters}, 106(19):190501, 2011.
	
	\bibitem{green2013arbitrary}
	Todd~J Green, Jarrah Sastrawan, Hermann Uys, and Michael~J Biercuk.
	\newblock Arbitrary quantum control of qubits in the presence of universal
	noise.
	\newblock {\em New Journal of Physics}, 15(9):095004, 2013.
	
	\bibitem{khaneja2005optimal}
	Navin Khaneja, Timo Reiss, Cindie Kehlet, Thomas Schulte-Herbr{\"u}ggen, and
	Steffen~J Glaser.
	\newblock Optimal control of coupled spin dynamics: design of nmr pulse
	sequences by gradient ascent algorithms.
	\newblock {\em Journal of magnetic resonance}, 172(2):296--305, 2005.
	
	\bibitem{warring2013individual}
	Ulrich Warring, C~Ospelkaus, Yves Colombe, R~J{\"o}rdens, Dietrich Leibfried,
	and David~J Wineland.
	\newblock Individual-ion addressing with microwave field gradients.
	\newblock {\em Physical review letters}, 110(17):173002, 2013.
	
	\bibitem{james1997quantum}
	Daniel~FV James.
	\newblock Quantum dynamics of cold trapped ions with application to quantum
	computation.
	\newblock Technical report, 1997.
	
	\bibitem{sorensen2000entanglement}
	Anders S{\o}rensen and Klaus M{\o}lmer.
	\newblock Entanglement and quantum computation with ions in thermal motion.
	\newblock {\em Physical Review A}, 62(2):022311, 2000.
	\bibitem{duan2024}
	Guo S A, Wu Y K, Ye J, et al. 
	\newblock A site-resolved two-dimensional quantum simulator with hundreds of trapped ions.
	\newblock {\em Nature}, 1-6, 2024. 

	\bibitem{allcock2021omg}
	DTC Allcock, WC~Campbell, J~Chiaverini, IL~Chuang, ER~Hudson, ID~Moore,
	A~Ransford, C~Roman, JM~Sage, and DJ~Wineland.
	\newblock Omg blueprint for trapped ion quantum computing with metastable
	states.
	\newblock {\em Applied Physics Letters}, 119(21):214002, 2021.
	
	\bibitem{singh2023mid}
	K~Singh, CE~Bradley, S~Anand, V~Ramesh, R~White, and H~Bernien.
	\newblock Mid-circuit correction of correlated phase errors using an array of
	spectator qubits.
	\newblock {\em Science}, 380(6647):eade5337, 2023.
	
	\bibitem{wootters1998entanglement}
	William~K Wootters.
	\newblock Entanglement of formation of an arbitrary state of two qubits.
	\newblock {\em Physical Review Letters}, 80(10):2245, 1998.
	
\end{thebibliography}
\end{document}